# A Time-constraint Satisfying and Cost-reducing node evaluation metric for Message Routing in Mobile Crowd Sensing Networks

Qian Wang, Zhipeng Gao, Kun Niu, Yang Yang, Xuesong Qiu
Beijing University of Posts and Telecommunications
Beijing, China
{wangqian1991, gaozhipeng, niukun, yyang, xsqiu}@bupt.edu.cn

## ABSTRACT
In mobile crowd sensing networks data forwarding through opportunistic contacts between participants. Data is replicated to encountered participants. For optimizing data delivery ratio and reducing redundant data a lot of data forwarding schemes, which selectively replicate data to encountered participants through node's data forwarding metric, are proposed. However most of them neglect a kind of redundant data whose Time-To-Live (TTL) is expired. For reducing this kind of redundant data we proposed a new method to evaluate node's data forwarding metric, which is used to measure the node's probability of forwarding data to destination within data's constraint time. The method is divided into two parts. The first is evaluating nodes whether have possibility to contact destination within time constraint based on transient cluster. We propose a method to detect node's transient cluster, which is based on node's contact rate. Only node, which has possibility to contact destination, has chances to the second step. It effectively reduces the computational complexity. The second is calculating data forwarding probability of node to destination according to individual ICT (inter contact time) distribution. Evaluation results show that our proposed transient cluster detection method is more simple and quick. And from two aspects of data delivery ratio and network overhead our approach outperforms other existing data forwarding approach.

## 1. INTRODUCTION
Mobile crowd sensing (MCS) aims to provide a mechanism, which employs participants from the publics to efficiently and effectively contribute sensing data from their mobile devices and utilize these data to solve specific problems in collaborations [1]. Thus data forwarding is a major process of mobile crowd sensing. Different from most of existing mobile sensing applications, which consider mobile users report and access sensory data through the Internet by cellular networks. Data forwarding among participants in mobile crowd sensing networks through opportunistic contacts with short-range radio communications.

Epidemic routing protocol [2] is the first and the most generic opportunistic data forwarding protocol. It optimizes forwarding performance by utilizing all contacted nodes for forwarding data. The carried data node forwards data through replicating data to the encountered node. Although its forwarding performance is optimized, it produces large amounts of redundant data. Thus a number of opportunistic data forwarding protocols have been proposed to reduce these redundant data, which replicate data to relays based on different data forwarding metrics. Node's forwarding metric is evaluated by predicting the node's capability of contacting others in the future. However, these opportunistic data forwarding protocols don't consider a kind of redundant data that is produced by data which has Time-To-Live (TTL). In the process of data forwarding, the data, which is carried by nodes that can't direct or indirect contact destination within data's time constraint, is redundant. For example, some applications based on mobile crowd sensing need to get sensing data which have time constraints, such as environment monitoring [3] [4], noise mapping [5], etc. Sensing data makes no sense if it doesn't deliver to the destination within time constraint.

Opportunistic data forwarding protocol consists of two parts: First, data forwarding metric: it is used to measures the node's probability of forwarding data to destination within the time constraint. Second, data forwarding strategy: a node replicates data to encountered node whose forwarding metric is higher than itself. In this paper, we propose a method to evaluate node's data forwarding metric, which considers data have time constraints. It reduces redundant data and at the same time forwarding performance is optimized. Evaluating data forwarding metric of node consists of two steps: First, evaluating contacted node's possibility of forwarding data to destination. Only node with high possibility has the chance to forward data. Second, calculating probability of the node, which has chance, forwards data to destination within data's time constraint.

Recently, some schemes [6][7][8] propose data forwarding metrics by exploiting the social contact patterns of human beings. [6] [8] observe that transient social contact patterns of mobile nodes during short time periods are usually different from their cumulative contact patterns. By aggregating contact information into a aggregated contact graph, important contact information such as the burst behavior will be lost, so it proposes transient community which normally appear during a time period and disappear thereafter. [7] observes that efficient content dissemination is mostly due to high contact rate nodes, and it shows that high contact rate users who spend more time in temporal communities, which are clusters of nodes that meet more frequently and for longer periods of times, are less efficient content forwarders.

Hence transient cluster detection based on pairwise contact rate, which indicates that data are transmitted quickly in cluster, and data transmission between clusters through bridging node, which is a node of cluster and also has strong connected to nodes of other clusters. Evaluating nodes whether have possibility to contact destination based on transient cluster. It effectively reduces the computational complexity.

Inter contact time (ICT) is defined as the time elapsing between two consecutive encounters of any two nodes. The distribution of ICT plays a key role in determining the performance of forward protocol. Because it expresses the probability that two nodes directly communicate within time constraint. It has recently been observed that the ICT distribution of nodes in real-world mobility traces displays some probability distribution models, which can be used to calculate probability of the node forwards data to destination within data's time constraint.

Most opportunistic data forwarding protocols use the aggregate distribution power law + exponential tail or exponential, which is obtained by considering the samples from all pairs together, to calculate probability of individual pair encountered. However, [9] proposes the distribution of aggregate ICT provides a correct representation of individual ICTs when all individual pairs distribution are independent and identically distribution (iid). However this is not correct in general if the network is heterogeneous. In the heterogeneous networks, individual pair ICTs can be exponential distribution, Pareto distribution, Log-normal distribution, etc. Therefore this should be taken in great care when using aggregate ICT as an indicator of the individual pair distribution.

The contributions of this paper are three-fold:

1). Every node has a transient cluster. We propose a method to detect node's transient cluster: comparing with existing transient community detect methods such as DRAFT (Distributed Rise and Fall spatio-Temporal) [10] and CCM (Contact-burst-based Clustering Method) [11], our method is more simple and quick.

2). We first judge node whether has possibility to contact destination based on transient cluster. If it does not have, the following steps would not be continued. This way effectively reduces the following computational complexity.

3). Calculating nodes' data forwarding probability to destination according to individual ICT distribution. Whether the encountered node has higher probability to destination considering the time constraint of the data. Simulation results show that our approach outperforms other existing data forwarding approach.

The rest of the paper is organized as follows. Section Ⅱ gives a brief overview of the related work. In section Ⅲ, we give a overview of our approach. Section Ⅳ presents transient community detection, section Ⅴ formulate the probability of node transmit data to destination, section Ⅵ evaluate the performance of our approach by simulation, and section Ⅶ concludes the paper.

## 2. RELATED WORK

Opportunistic data forwarding protocol originates from Epidemic routing [2] which floods the network. Later studies develop forwarding protocols to approach the performance of Epidemic routing with lower cost, which is measured by the number of data copies in the network. Quota-based routing protocols [12] [13] achieve more reasonable delivery cost by restricting the number of data copies. Specifically, at any point in time, a fixed number of each data exists in the whole network. However, the one size fits all approach for data forwarding is insufficient. Thus most data forwarding protocols do not limit the number of data copies and forward data by comparing the nodes' data forward metrics.

Data forwarding metric measures the nodes' capability of forwarding data to the destination. Methods are used to estimate node's data forwarding metric can be divided into two categories. The first category, node mobility pattern is exploited based on contact history and then be used to predict the nodes' capability of contacting others in the future, which is node's forwarding metric. Such as PRoPHETv2 [14] uses the history of previous encounters to estimate delivery predictability for data, and MaxProp [15] estimates the node contact likelihood based on the contact counts in the past. However, in all methods mentioned above, node contact capability is predicted only based on the direct contacts among mobile nodes. They don't consider indirect contacts among nodes, which contribute to improving data forwarding performance.

The second category, node's forwarding metric is estimated based on communities, which formed by exploiting nodes' social contact patterns. The utilization of community structure has been proved to improve routing performance [16]. Therefore, the community structure has been widely utilized in the design of routing protocols. In SimBet [17] and BUBBLE Rap [16], betweenness [18] is used as the forwarding metric, which measures the social importance of a node facilitating the communication among other nodes. However, node's forwarding metrics in these protocols are evaluated based on the previous cumulative network knowledge. Transient characteristics of node contact pattern, which influence data forwarding performance to a large extent, are ignored.

Thus, [10] [19] [20] [21] are proposed considering transient connectivity and indirect contacts among nodes. [10] aims to improve the relevance of clusters, which are formed by participants who meet most often, to particular time frames. And thus improve the performance of cluster based data delivery algorithms. However this data forward algorithm doesn't consider the node, which is determined to take duplicated data, whether can contact destination through direct or indirect method within data's time constraint. Therefore it maybe produces some unnecessary redundant data.

[19] proposes a novel approach which using both k-clique and modularity community detection method detect the transient community structure in both traces every hour. And then node's data forwarding metric is based on the node's centrality, which is measured by using the transient contact distribution and transient community within the given scope and time constraint. Although the proposed forwarding metrics achieve much better performance compared to some existing schemes with similar forwarding cost, the transient community structure detect method is complex. And node's data forwarding metric based on centrality, which is measured only in its transient community. It doesn't reflect the nodes' capability of forwarding data to a specific destination. Thus this data forwarding metric doesn't suit mobile crowd sensing network opportunistic data forwarding protocol.

[20] presents a novel adaptive multi-step protocol. In each routing step this protocol reasons on the remaining time-to-live of the message in order to allocate the minimum number of copies

necessary to achieve a given delivery. Because the source of a message uses the community topology to pre-compute the multi-hop path that traverses the minimal number of communities through their gateway nodes and that has the highest delivery probability. This method is impractical sometimes because we cannot get topology of all communities in some cases.

[21] proposes an expected encounter based routing protocol (EER) which distributes multiple replicas of a message proportionally between two encounters according to their expected encounter value. It only considers direct contact and ignores indirect contact. At the same time it proposes a community aware routing protocol (CAR), in which the distribution of the replicas of this message is proportional to the expected numbers of encountering communities of each pair of encounter. This method is impractical sometimes because we cannot get the expected numbers of encountering communities in advance in some cases.

In this paper, we consider direct, indirect contacts among nodes. And transient characteristic of nodes' contact pattern influences data forwarding performance. Therefore we propose a new method to evaluate node's data forwarding metric, which aims to improve mobile crowd sensing network's data forwarding performance. This method is divided into two steps: first, we propose a method to detect node's transient cluster, which varies based on real time condition. It is simple and quick comparing to existing transient community detection method. Judging contacted node's possibility of forwarding data to destination based on node's transient clusters. Second, calculating probability of the node, which has possibility, forwards data to destination within data's time constraint based on individual pair ICT.

## 3. SYSTEM MODEL

Opportunistic data forwarding protocol consists of data forwarding metric and data forwarding strategy. When the carried data node encounters another node. Data forwarding mechanism is as following.

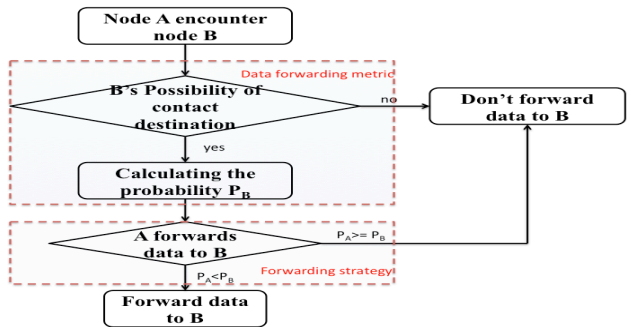

**Figure 1. Data forwarding mechanism**

First, calculating the encountered node's data forwarding metric. Data forwarding metric is divided into two steps. First step, judging whether the encountered node has possibility of contact destination within time constraint of data according to node's transient cluster which demonstrates the time-varying relationship of nodes' contact. Only node with possibility has the chance to forward data. Second step, using ICT distribution between nodes to calculate probability of the node, which has chance, forwards data to destination within data's time constraint.

Second, determining whether forwarding data to the encountered node according to forwarding strategy. A node forwards data to the encountered node whose forwarding metric is higher than it.

Opportunistic data forwarding mechanism is illustrated in Figure1. In this paper, we focus on evaluating node's data forwarding metric.

## 4. EVALUATIONG DATA FORWARDING METRIC OF NODES

Evaluating node's data forwarding metric consists of two parts. First part is judging the contacted node whether has possibility to encounter destination based on its transient cluster. Second part is calculating the probability.

### 4.1 Transient cluster detection

Every node has a transient cluster, which is different from some communities that reflect nodes' social relationship. Node's transient cluster is formed opportunistically by pair-wise nodes' encounter rate and reflects the set of nodes, which have higher contact rate with it. [11][19] propose pair-wise contacts show the burst behavior，which is different from cumulative contact. So node's transient cluster is time-varying and it would delete the node whose contact rate is lower than pre-defined threshold. A transient cluster is a dynamic encounter graph.

Detecting node's transient cluster needs parameter x, which is a pre-defined threshold. And the value of x is set based on different traces. Node belongs to a node's transient cluster if and only if the inter-contact time between them is shorter than x, or the node is deleted from the transient cluster it belongs.

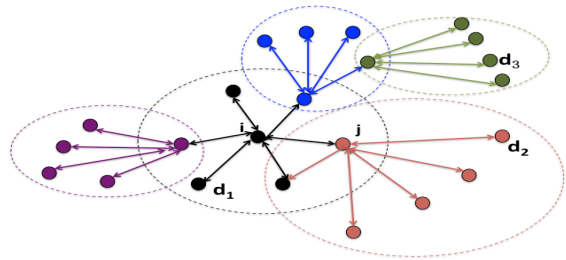

**Figure 2. Node's transient cluster and adjacent clusters**

[22] observes that efficient content dissemination is mostly due to high contact rate nodes. Thus node's transient clusters can be used as judgment whether forwarding data to it or not. Because [23] proposes limiting a node's neighborhood view to four hops is enough to improve forwarding efficiency while keeping control overhead low for the dataset they consider. [10] proposes they don't check for nodes further than 2 hops away as this would require nodes exchange and store a large amount of additional data. Thus, through comprehensive consideration the encountered node has possibility to contact destination, only when its transient cluster or adjacent clusters contain destination. For example, the carried data node n encounters node $i$. At that time $i$'s transient cluster and its adjacent clusters were illustrated in Figure 2. If the data's destination is node $d_1$ or $d_2$, which is a member of node $i$'s transient cluster or adjacent clusters. Node $i$ would have possibility to encounter destination. But if data's destination is node $d_3$, node $i$ would not have possibility to encounter destination.

Transient clusters of nodes change over time, and data has time constraint $t_D$. Thus the relationship between them would influence the accuracy of the decision. If node's transient cluster's duration $t_C > t_D$, the decision would be the most appropriate. Or it is not the best. Because it is possible that the node's current transient cluster doesn't contain destination, but it contains destination when it changes. And at the same time the time doesn't exceed data's valid time. A node $d_i$ maintains the following information. A set of nodes belongs to $d_i$'s transient cluster $T_i$. And Node $d_j$'s transient cluster members, $d_j$ belongs to $T_i$.

The process with which $d_i$'s transient cluster $T_i$ is built can be summarized as following. This process doesn't need centralized control. It performs independently and real-time.

1) Initially $T_i$ is set to $\{\Phi\}$

2) When $d_i$ encounters another node $d_j$, if the inter-contact time, which defined as the time elapse between two successive contact period for a given pair of nodes, is shorter than x, $d_j$ is added to $d_i$'s transient community $T_i$ and the encounter time is recorded. After that, if the inter-contact time between them is bigger than x, $d_j$ is deleted from $d_i$'s transient community $T_i$.

3) Node $d_j$ belongs to $T_i$, node $d_j$ sends its information of transient cluster $T_j$ to node $d_i$, if $T_j$ changed, $d_j$ send the changed $T_j$ to node $d_i$.

## 4.2 Calculating the probability

Node has been determined that has the possibility to encounter destination after first step of node's data forwarding metric. Then we calculate the probability of the node encounters destination within data's time constraint, which is used as judgment standard of data forwarding strategy whether replicating data to the encountered node or not.

Node's transient cluster contains nodes, which have high contact rate with it. There is no guarantee that node would deliver data to its transient cluster or adjacent clusters members immediately, even though paths between them at most 2 hops. It only means that the node has higher possibility to encounter destination within transient clusters duration. Thus for reducing network overhead and increasing data delivery ratio, in the paper we calculate the probability that the encountered node encounters destination within data's time constraint based on node's transient cluster. Thus the relationship between node's transient cluster time and data's valid time would influence time chooses of calculating the probability of node encounters destination. And data is replicated to the encountered node, only when its probability is higher than the carried data node.

Node's transient cluster is changing, and [19] proposes that the distribution of nodes' contact duration $f_c$ can be accurately approximated by normal distribution. Therefore the relationship between transient cluster duration time and data's valid time is divided into two possibilities that cluster's duration time is longer or shorter than data's valid time. It is illustrated in Figure 3. $t_e$ is the time that the carried data node encounter another node. $t_v$ is data's valid time. $t_{cs}$ and $t_{ct}$ is the start time and end time that encountered node and destination in same transient cluster.

① is the case that node's transient cluster time is shorter than data's valid time. Thus when calculating the probability of node encounters destination time is node's transient cluster's duration. ② is the case that node's transient cluster time is longer than data's valid time. In this situation calculating the probability of node encounters destination time is data's valid time.

ICT distribution of nodes in real-world mobility traces displays some probability distribution models. And [9] proposes in the heterogeneous networks ICT distribution of individual pairs can be exponential distribution, Pareto distribution, Log-normal distribution, etc. The distribution can be got from history of pairs' contact. In the paper, we calculate the probability of node encounter destination within data's time constraint using individual pairs' ICT distribution $f$ instead of aggregate ICT distribution.

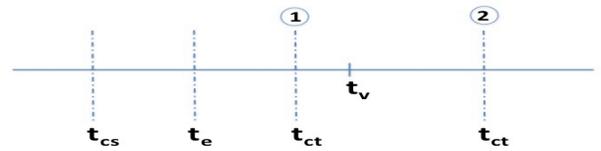

Figure 3. Relationship between cluster duration

and data's valid time

When a carried data node n encounters a node $i$, which is not destination. And node $i$ has been determined that has the possibility to encounter destination after first step of node's data forwarding metric. Calculating the probability of node $i$ encounters destination d within data's time constraint. It is divided into two situations.

1) Destination in node $i$'s transient cluster

2) Destination in node $i$'s adjacent transient clusters

## 4.2.1 Destination in node's transient cluster

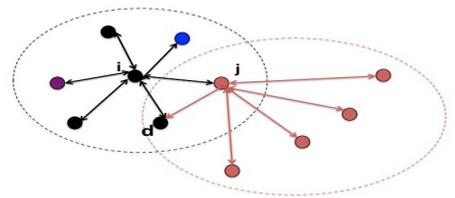

Figure 4. Destination in node's transient cluster

We consider data forwarding through direct contact and two hops indirect contact among nodes. Therefore, node $i$ contacts destination $d$ may exist two possibilities. It is illustrated in Figure 4. Node $i$ directly contacts destination $d$ or indirectly contacts destination $d$ through node $j$.

Within data's time constraint $(t_1,t_2)$ the probability of node $i$ encounters destination $P_c(t_1,t_2)$ is calculated as following. It contains direct and indirect contact. Equation (1) indicates the probability that node $i$ direct contact destination $d$. $f_{id}$ is the ICT distribution between node $i$ and node $d$. Equation (2) indicates the probability that node $i$ through relay $j$ contact destination d. Equation (3) is the probability that node $i$ through relays contact destination $d$. D contains nodes that contain node $d$ in their transient clusters. And they are in $i$'s transient cluster.

$$P_{id}^{(1)} = \int_{t_1}^{t_2} f_{id}\, d_t \tag{1}$$

$$P_{ijd} = \int_{t_1}^{t_2}(\int_{t_1}^{t_0} f_{ij}\, d_t \cdot \int_{t_0}^{t_2} f_{jd}\, d_t)\, d_{t_0} \tag{2}$$

$$P_{id}^{(2)} = \sum_{j \in D} P_{ijd} \tag{3}$$

$$P_c(t_1,t_2) = P_{id}^{(1)} + P_{id}^{(2)} \tag{4}$$

Data's valid time is $(t_s,t_v)$, and transient cluster's duration time is $(t_{cs},t_{ct})$. Because the relationship between the time, which encountered node and destination in same transient cluster, and data's valid time exists two possibilities. The time, which is used to calculate probability, is different in these two situations.

When $t_{ct} \geq t_v$, the probability $P_c^{(1)}(t_s,t_v)$ that node $i$ encounters destination $d$.

$$P_c^{(1)}(t_s,t_v) = \int_{t_v-t_{cs}}^{\infty} f_c \cdot P_c(t_e,t_v)dt \tag{5}$$

When $t_{ct} < t_v$, the probability $P_c^{(2)}(t_s,t_v)$ that node $i$ encounters destination $d$.

$$P_c^{(2)}(t_s,t_v) = \int_{t_e-t_{cs}}^{t_v-t_{cs}} f_c \cdot P_c(t_e,t_{ct})dt \tag{6}$$

In summary, equation (7) is the probability that node $i$ encounters destination within data's time constraint $(t_s,t_v)$.

$$P_c(t_s,t_v) = P_c^{(1)}(t_s,t_v) + P_c^{(2)}(t_s,t_v) \tag{7}$$

## 4.3 Destination in node's adjacent transient cluster

The situation is illustrated in Figure 5. Destination d is not in node $i$'s transient cluster. But it is in $i$'s adjacent transient clusters. Thus node $i$ contacts destination d through node j. Within data's time constraint $(t_1,t_2)$ the probability of node $i$ encounters destination d $P_c(t_1,t_2)$ is calculated as following. Equation (8) indicates the probability that node $i$ encounters node j within time constraint $(t_1,t_2)$. Equation (9) is the probability that node $i$ encounters node d through relays within time constraint $(t_1,t_2)$. T contains nodes whose transient clusters contain destination d. And they are in node $i$'s transient clusters.

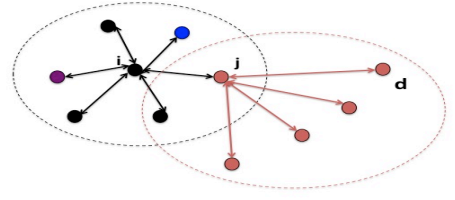

**Figure 5. Destination in node's adjacent cluster**

$$P_{ij}(t_1,t_2) = \int_{t_1}^{t_2} f_{ij}dt \tag{8}$$

$$P_{id}(t_1,t_2) = \sum_{j \in T}(\int_{t_1}^{t_2}(P_{ij}(t_1,t_3) \cdot P_{jd}(t_3,t_2))dt_3) \tag{9}$$

Data's valid time is $(t_s,t_v)$, and transient cluster's duration time is $(t_{cs},t_{ct})$. Because the relationship between the time, which encountered node and destination in same transient cluster, and data's valid time exists two possibilities. Time, which is used to calculate probability, is different in these two situations.

When $t_{ct} \geq t_v$, the probability $P_{id}^{(1)}(t_s,t_v)$ that node $i$ encounters destination $d$.

$$P_{id}^{(1)}(t_s,t_v) = \int_{t_v-t_{cs}}^{\infty} f_c \cdot P_{id}(t_e,t_{ct})dt \tag{10}$$

When $t_{ct} < t_v$, the probability $P_{id}^{(2)}(t_s,t_v)$ that node $i$ encounters destination $d$.

$$P_{id}^{(2)}(t_s,t_v) = \int_{t_e-t_{cs}}^{t_v-t_{cs}} f_c \cdot P_{id}(t_e,t_{ct})dt \tag{11}$$

Equation (12) is the probability of node $i$ encounters destination d within data's time constraint $(t_s,t_v)$.

$$P_{id}(t_s,t_v) = P_{id}^{(1)}(t_s,t_v) + P_{id}^{(2)}(t_s,t_v) \tag{12}$$

In summary, first, judging the destination's location that destination in encountered node's transient cluster or adjacent clusters. And then choosing appropriate formulations to calculate the probability that contacted node encounters destination within data's valid time.

## 5. PERFORMANCE EVALUATION

In this section, first we compare our transient cluster detection method with other cluster detection methods. Then we evaluate our proposed opportunistic data forwarding protocol from aspects of data delivery ratio and network overhead. Data delivery ratio is proportion of data items successfully delivered before data expires. The network overhead is the number of data copies created in the network.

Our experiments are based on three real date sets, MIT Reality [24], Infocom6 [16] and Cambridge [25]. These data is formed based on the devices periodically detect their peers via Bluetooth interfaces, and a contact is recorded when two devices move into the communication range of each other. The details of three data sets are shown in Table 1.

## Table 1 traces summary

|  | Infocom6 | Cambridge | MIT Reality |
|---|---|---|---|
| Environment | Conference | City | Campus |
| Duration (days) | 3 | 12 | 246 |
| Number of devices | 78 | 36 | 97 |
| Inter-probe time(s) | 120 | 600 | 120 |

## 5.1 Transient cluster evaluation

We compare our proposed transient cluster detection (TCD) method with similar cluster detection methods: Contact-burst-based Clustering Method (CCM) [11] and Distributed Rise and Fall spatio-Temporal (DRAFT) clustering method [10].

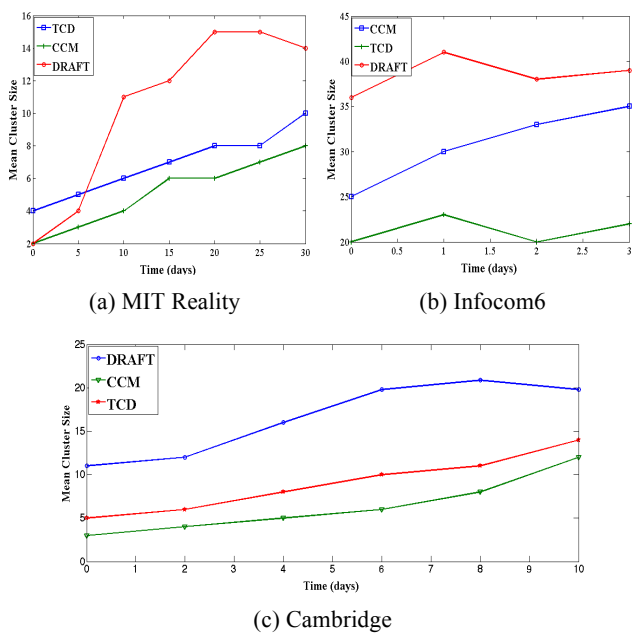

(a) MIT Reality  (b) Infocom6

(c) Cambridge

**Figure 6. Mean cluster size**

**CCM cluster detection method:** Detecting transient clusters by exploiting the pairwise contact processes. It is based on the detailed pairwise contact information between nodes. CCM formulates each pairwise contact process as regular appearance of contact bursts, during which most contacts between the pair of nodes appear. Based on such formulation, detecting transient clusters by clustering the pairs of nodes with similar contact bursts together.

**DRAFT cluster detection method:** it provides spatio-temporal, non-social clustering within dynamic encounter graphs. It combines spatial clustering with a decay function. And clusters reflect current and recent behavior patterns by excluding devices, which have not been seen for a long time. It uses three parameters to govern the rate at which clusters grow and decay. Nodes are added in or deleted from clusters according to cumulative time.

The performance is compared based on three metrics: mean cluster size, max cluster size and mean 2-hop clusters size. The value of parameter x, which is used to judge whether two nodes belong to same transient or not, is set empirically based on different traces. In these three data sets we found that with x= 1 hour most contacts happen within some contact bursts which only account for small portion of total time. And the performance is not very sensitive to the change of x according to experiments. Thus we set x=1 hour.

**Mean cluster size:** Figure 6 shows that clusters detected by TCD is smaller than those detected by DRAFT and CCM. At same time our proposed cluster detection method is more simple and rapid than DRAFT and CCM. Because cluster detected by TCD method reflects nodes, which have high contact rate with a node. However cluster detected by CCM reflects nodes whose contact burst with similar duration. And in DRAFT method node is added into or deleted from a cluster based on the cumulative or decayed encounter duration time.

**Change of cluster size**: we randomly chose one day and a cluster. We recorded the change of cluster size in this day, which is illustrated in Figure 7. In this part, we compare TCD and DRATF. Because only these method are real-time changed. The change of cluster based on TCD detection method is more frequent than DRAFT method. Because if encounters are frequently interrupted by lost neighbor discovery request, then basing cluster membership on single encounter durations or the time between encounters will be unreliable. Thus in DRAFT method clusters are changed based on cumulative or decayed time. However for some cases that clusters' size change frequently and data's valid time is short. The method, which can real-time reflect nodes' contact rate, performs better in data delivery performance. TCD cluster detection method is stronger in real-time comparing to DRAFT and CCM method. Therefore in some cases TCD method is more suitable.

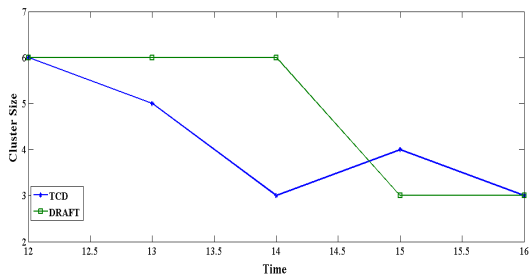

**Figure 7. Change of cluster size**

Figure 8 shows that max cluster, which is detected by DRAFT contains more than half of the whole nodes. Thus for some large experiment this method may not be scalable. Although max cluster detected by CCM and TCD method is smaller than which is detected by DRAFT. Max cluster detected by CCM is bigger than our proposed TCD method.

DRAFT and our proposed method judge whether replicate data to the encountered node based on this node's 2-hop nodes. Thus the size of node's 2-hop nodes is very important for these methods' data forwarding performance. Mean 2-hop clusters size in Figure 9 (a) shows that in the Infocom6 the number of 2-hop nodes detected by DRAFT is on average 85% larger than the number of nodes in local cluster. Figure 9 (a) (c) show that in MIT Reality and Cambridge data sets the number of 2-hop nodes detected by DRAFT is much larger, with on average two times more than nodes in local cluster. In conclusion comparing with DRAFT method our proposed method detected the number of 2-hop nodes is much smaller, which is more useful for data forwarding.

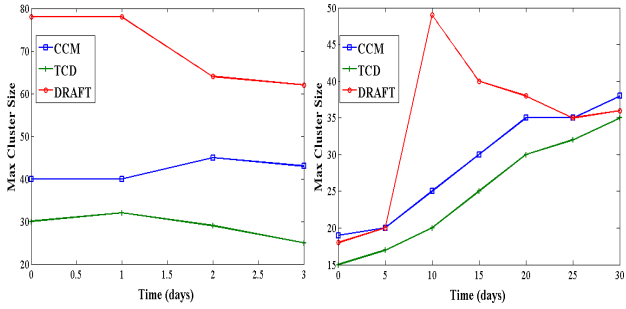

(a) Infocom6  (b) MIT Reality

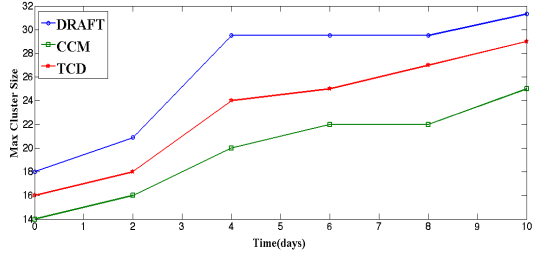

(c) Cambridge

**Figure 8. Max cluster size**

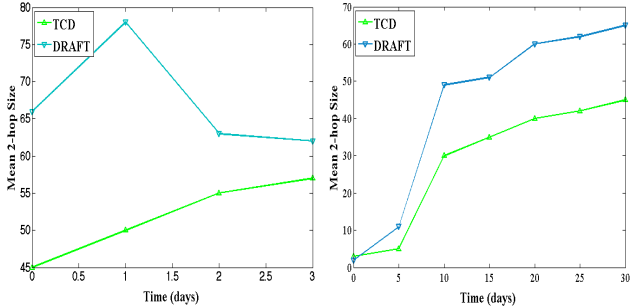

(a) Infocom6  (b) MIT Reality

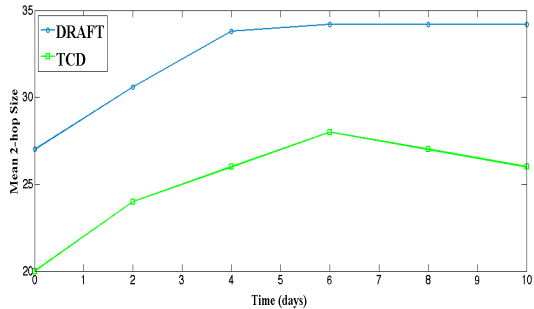

(c) Cambridge

**Figure 9. Mean 2-hop size**

## 5.2 Data forwarding performance evaluation

In the experiment for fairness sources and destinations are picked randomly and the data's generation time is randomly chosen in the daytime, since nodes' activity time is low at night which may results in inaccurate comparison. The performance is measured with two metrics: one is data delivery ratio and the other is network overhead. Our proposed TCD method is compared with a traditional community forwarding approach Bubble Rap, Epidemic approach, and two similar methods DRAFT, CCM.

**Bubble Rap**: this strategy uses both centrality and community. CPM (K-clique) is used to detect communities. The data item is always forwarded to a higher centrality node, until it reaches a node that belongs to the same community as the destination node. When the data item reaches the destination community, it is forwarded to higher-centrality node within the community's scope, until the destination node is reached.

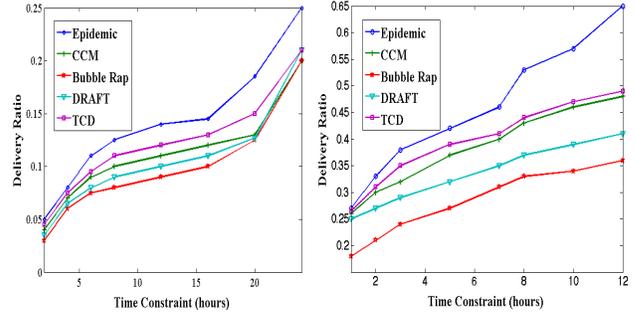

(a) MIT Reality  (b) Infocom6

**Figure 10. Data delivery ratio**

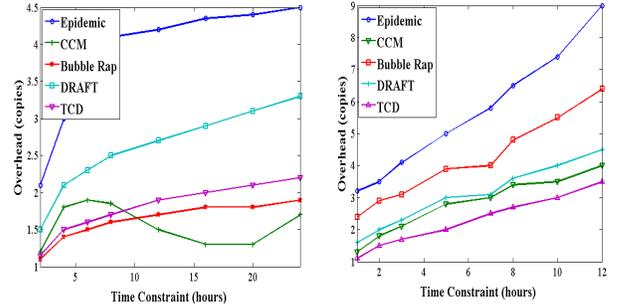

(a) MIT Reality  (b) Infocom6

**Figure 11. Network Overhead**

The results are shown in Figure 10 and Figure 11. Generally speaking, our proposed method achieves higher data delivery ratio. When the time constraint T is short, the data forwarding performance of cluster detection methods, which reflect nodes' transient characteristics is better than those reflect nodes' cumulative characteristics. Because nodes' cumulative characteristic could not reflect the character that node's contact rate changes over time. Especially when T< 3 hours, our proposed method and CCM achieve similar delivery ratio with that of Epidemic. With the increase of T the advantage of transient cluster detection comparing to cumulative cluster detection method is decreased.

For our proposed method we notice that with the increase of T comparing with Epidemic the advantage of our proposed method decrease. The reason is we neglect some nodes, which don't contain destination node currently. Although these nodes wouldn't influence data delivery performance when time constraint T is small. With the increase of data's time constraint T these nodes have higher possibility to contact destination within data's valid time. Neglecting this possibility would influence data forwarding performance. Figure 11 show that the forwarding cost of our proposed method is lower than other methods. Because we calculate probability that node encounter destination within data's time constraint. Data would not be replicated to nodes, which have lower probability. Thus network overhead are decreased.

In conclusion, the data forwarding performance of our proposed method is better than Bubble Rap, Epidemic approach, DRAFT. Although the performance is similar, our proposed method is simple, rapid and real-time comparing to CCM.

## 6. CONCLUSION

In this paper, we propose a method to evaluate node's data forwarding metric. It is divided into two steps. For reducing computational complexity, we first judge whether the encountered node has possibility of forwarding data to destination based on our proposed transient cluster detection method. Node's transient cluster reflects nodes, which have high contact rate with it. And it changes according to the real condition. Second using ICT distribution between nodes to calculate probability of the node, which has chance, forwards data to destination within data's time constraint. Only the node's forwarding metric is higher than the carried data node. It has opportunity to carry data. Simulation results show that our approach outperforms other existing data forwarding approach.

## ACHNOWLEDGMENTS
This research is supported by NSFC (61272515, 61372108, 61401033) and National Science & Technology Pillar Program (2015BAH03F02).